\documentclass[superscriptaddress,aps,prl,floats,twocolumn,twoside,floatfix]{revtex4}
\usepackage{epsfig}
\usepackage{amsmath}
\usepackage{amssymb}

\newcommand{\be}{\begin{equation}}
\newcommand{\ee}{\end{equation}}
\newcommand{\bea}{\begin{eqnarray}}
\newcommand{\eea}{\end{eqnarray}}
\begin{document}
\title{Comment on ``Both site and link overlap distributions are non trivial in 3-dimensional
Ising spin glasses'', cond-mat/0608535v2}  
\author{Pierluigi Contucci}
\affiliation{
Universit\`{a} di Bologna, Piazza di Porta S.Donato 5, 40127 Bologna, Italy}
\author{Cristian Giardin\`a}
\affiliation{Eurandom, P.O. Box 513 - 5600 MB Eindhoven, The Netherlands}
\maketitle
In a recent interesting paper G.Hed and E.Domany
\cite{HD}
have investigated the properties of 3-dimensional
Ising spin glasses by numerical methods and found that link and site overlap distributions are both
non trivial.
They moreover claim that although the two overlap distributions are non trivial their behavior
is not well described by mean field theory as predicted by the Replica Symmetry Breaking
solution \cite{MPV}. Their argument is based on the disproof, by numerical tests, of one of the simplest
link-overlap identities built in the Parisi theory of the spin glass phase:
\begin{equation}\label{ide}
p(Q_{12},Q_{34})=\frac{2}{3}p(Q_{12})p(Q_{34})+\frac{1}{3}p(Q_{12})\delta(Q_{12}-Q_{34})\; .
\end{equation}
The violation of the same identity for the site overlap was already claimed in a previous paper
\cite{HYD}.

Being such an identity false for the 3-dimensional Edwards-Anderson model the
Parisi mean field theory would turn out to be not correct
for the description of the low temperature phase. Within the language of RSB the previous
statement is explained by the mutual relation of two algebraic properties of the overlap matrix:
{\it ultrametricity} and {\it replica equivalence}. A symmetric matrix with a hierarchical block
structure (ultrametricity) fulfill indeed the property that each line is a permutation of any other line
(replica equivalence). See \cite{P1} for a detailed account on the two ansatz.
The algebraic property of replica equivalence can be translated into
a property of the multi-overlap probability distributions for real replicas (identical copies of the system
correlated by the same disorder) like equation (\ref{ide}) or similar identities.
See \cite{Gume, AC, GG} for a rigorous account in the mean
field Sherrington-Kirkpatrick model and \cite{P1}
for the equivalence between the algebraic and probabilistic description.

The aim of this short note is twofold: first to clarify what replica equivalence claims in its probabilistic version
and in particular to stress the fact that it has been \underline{rigorously proved} not only for the SK
model but also for the Edwards-Anderson model in terms of the link overlap \cite{C,ANN,CG};
second to propose a different interpretation of (part of) the numerical results in \cite{HD}.

The proof of replica equivalence  in finite-dimensional short-range spin glasses is based on two
ingredients: the first is the invariance under perturbations
of the quenched spin glass measure ({\it stochastic stability}). The second is the control
of the fluctuations of the internal energy ({\it self averaging}). The final result says that suitable
combinations of link overlap moments integrated over arbitrary temperature intervals
must have zero value when the thermodynamic limit is reached. Since the variables
under consideration are bounded the result is immediately extended to the distribution
functions and imply for instance that the identity (\ref{ide}) hold integrated over any {\it arbitrary}
$\beta$ intervals.

The result of the paper \cite{HD} concerning the violation of one identity is obtained
on a numerical test at a fixed temperature and shows that for the volumes under investigation,
cubes of side from $L=8$ to $L=12$, the finite volume distributions do not seem to converge
to equation (\ref{ide}).

In view of the fact that the mentioned identities have been rigorously proved for
the link overlap, we believe that the observations made in \cite{HD} can be interpreted as
a non monotonicity (for small volumes) of the tendency toward those identities.
Moreover since it has been numerically well established the relative non fluctuation
of the two overlaps in \cite{CGGV} (consistently with \cite{HD}) we believe that also the
results in \cite{HYD} concerning the violation of replica equivalence for the site overlap
could be interpreted in the same way.
One has also to notice from previous numerical studies on site overlaps moments \cite{MPRRZ} that,
on finite systems sizes, violations of the site overlap identities are very small (${\cal O}(10^{-4})$).
Due to large statistical errors in the low temperature phase it is difficult in general to
observe a monotonic behavior towards zero as the system size is increased.

In conclusion although there isn't yet an agreement on the nature of the
low temperature phase for the 3D spin glass model we believe that all
the interpretations of the numerical results which aim to disproof
the replica equivalence property for the link overlap must be rejected.

{\bf Remark.} See also the successive version cond-mat/0608535v4

{\bf Acknowledgments.} We want to thank G.Hed and E.Domany for a useful correspondence.

\end{document}